# Integrated Access Backhauled Networks

Oumer Teyeb, Ajmal Muhammad, Gunnar Mildh, Erik Dahlman, Filip Barac, Behrooz Makki
Ericsson, Ericsson Research, Sweden
{oumer.teyeb, ajmal.muhammad, gunnar.mildh, erik.dahlman, filip.barac, behrooz.makki}@ericsson.com

*Abstract*—5G is finally here. Initial deployments are already operational in several major cities and first 5G-capable devices are being released. Though it is not limited only to millimeter wave deployments, the main promise of 5G lies in the utilization of the high bandwidth available at high frequencies. However, high-frequency deployments are coverage-limited and require denser placement of base stations, which can increase the cost significantly. One of the main contributing factors to the cost is fiber deployment. Integrated access backhauling (IAB), where part of the wireless spectrum is used for the backhaul connection of base stations instead of fiber, is an attractive solution that could make dense deployments economically viable. With this main objective, 3GPP is in the process of standardizing multi-hop IAB networks. This paper provides an overview of the main features of the multi-hop IAB 3GPP rel-16 standard and the rationale behind the design choices.

## I. INTRODUCTION

According to latest reports, global mobile data traffic grew by 82% only in one year (between Q1 2018 and Q1 2019), mainly driven by the rising number of smartphone subscriptions and an increasing average data volume per subscription, most of which is video content [1]. It is estimated that the growth will continue, with an annual rate of 30%, between 2018 and 2024.

Meanwhile, the first 5G deployments are already operational. During the first half of 2019, several operators started offering 5G services and more and more smartphone vendors are releasing 5G-compatible handsets. One of the salient features of 5G is the usage of high millimeter wave (mmWave) frequencies, mainly due to the availability of large spectrum at these bands. However, high-frequency deployments are coverage-limited and require denser placement of base stations.

The main challenge in network densification is the site acquisition costs and fiber deployment. For example, a fiber-optic link is estimated to cost 100,000−200,000 USD/km in metropolitan areas, with a large portion (85%) of the total figure tied to trenching and installation [2]. For this reason, wireless backhauling is an attractive replacement for fiber, providing almost the same rate as fiber optic, but with significantly lower cost. Consequently, integrated access and backhaul (IAB) networks, where part of the radio resources is also used for wireless backhauling, has recently received considerable attention [3], [4].

IAB has been studied earlier in 3GPP in the scope of LTE Rel-10, also known as LTE relaying [5]. However, there have been only a handful of commercial LTE relay deployments, mainly because the existing LTE spectrum is too expensive to be used for backhauling, and small-cell deployments did not reach the anticipated potential in the 4G timeline.

For 5G new radio (NR), IAB is currently being standardized for 3GPP rel-16, which is expected to be completed by Q1 2020 [6]. The main reason why NR IAB is expected to be more commercially successful than LTE relaying, is the fact that the short-range of mmWave access creates a high demand for densified deployments, which, in turn, increases the need for backhauling. Moreover, the larger bandwidth available in mmWave spectrum provides more financially viable opportunity for self-backhauling. Additionally, multi-beam systems and MIMO, which are inherent features of NR, reduce cross-link interference between backhaul and access links allowing higher densification.

In this paper, we provide an overview of the multi-hop IAB system that is currently being standardized by 3GPP and explain the rationale behind the design choices. Section II gives an overview of the NR architecture. In Section III, the key design objectives and features of IAB networks in 3GPP are provided. Section IV gives the architecture and higher-layer protocol overview of multi-hop IAB networks, while Section V describes the physical layer aspects. Finally, concluding remarks are given in Section VI.

## II. OVERIVEW OF NR ARCHTIECTURE

Fig. 1 illustrates the architecture of a 5G network. The gNB is a base station providing NR user plane (UP) and control plane (CP) protocol terminations towards the user equipment (UE) and is connected via the NG interface to the 5G core network (5GC). The ng-eNB is an LTE base station that is connected to a 5GC. The user plane function (UPF) is responsible for functions related to the handling of UP data, such as packet routing, QoS handling and being an anchor point for mobility. The access and mobility management function (AMF) is responsible for CP functions such as access security/authentication/authorization control, paging, and mobility management.

The gNB can be one single logical node or it may consist of a central unit (CU) and one or more distributed unit(s) (DU(s)). A CU is a logical node hosting radio resource control (RRC), service data adaptation protocol (SDAP) and packet data convergence protocol (PDCP) of the gNB that controls the operation of one or more DUs.

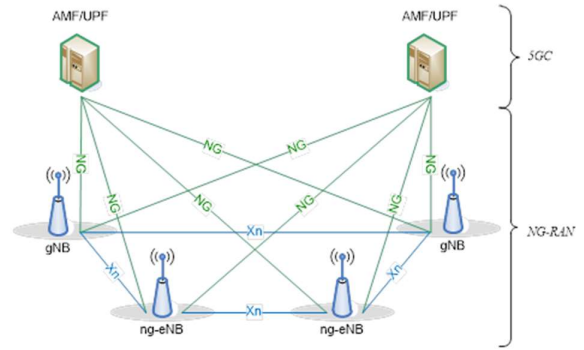

Fig. 1. Overall 5G architecture.

A DU is a logical node hosting radio link control (RLC), medium access control (MAC) and physical (PHY) layers of the gNB. The CU and the DU(s) it controls are connected via the F1 interface. The F1 application protocol (F1-AP) is used for conveying the lower-layer configuration information of the radio bearers between the CU and DU, as well as the setting up of a GTP tunnel between the DU and CU for each radio bearer.

The motivation for the CU/DU functional split is that all time-critical functionalities such as scheduling, fast retransmission, segmentation etc. can be realized in the DU (i.e. close to the radio and the antenna), while it is possible to centralize and resource-pool the less time-critical radio functionalities in the CU, even if non-ideal backhaul is used between the CU and DU. Fig. 2 illustrates the CU/DU split architecture.

An additional benefit of the CU/DU separation is that all external interfaces of the gNB, such as Xn (i.e. the interface between neighboring gNBs) are terminated in the CU, thus avoiding the extra complexity of terminating external interfaces to every DU. The CU/DU split also supports centralized termination of the PDCP, which facilitates both the security protection of UE communication end-to-end (E2E) between the UE and the CU, as well as the packet handling at dual connectivity and handover, since the traffic flows to/from different DUs are separated at the CU.

## III. KEY DESIGN OBJECTIVES AND FEATURES

A key benefit of IAB is enabling flexible and very dense deployments of NR cells without the need to densify the transport network proportionately. A diverse range of deployment scenarios can be envisioned, including support for outdoor small cell deployments, coverage extension, indoor deployments and fixed wireless access (FWA). This is illustrated in Fig. 3.

IAB work has been ongoing in 3GPP since 2017 and several design approaches were discussed, with the main criteria considered being the effective and flexible deployment of a system that allows a smooth transition and flexible integration from/to legacy deployments.

The key features to be supported by the first release of 3GPP IAB network for NR backhauling (rel-16), which is expected to be completed by Q1 2020, are:

- **Multi-hop backhauling:** to enable flexible range extension.
- **QoS differentiation and enforcement:** to ensure that the 5G QoS of bearers is fulfilled even in a multi-hop setting.
- **Support for network topology adaptation and redundant connectivity:** for optimal backhaul performance and fast adaptation to backhaul radio link overloads and failures.
- **In-band and out-of-band relaying:** the use of the same or different carrier frequency for the access (i.e. link to UEs) and backhaul links (i.e. link to other network nodes) of the IAB node, respectively.
- **Support for legacy terminals:** the deployment of IAB nodes should be transparent to UEs (i.e. no new UE features/standardization required).

## IV. ARCHITECTURE PRINCIPLES AND PROTOCOL ASPECTS

### A. General Overview

The simplest way to design a multi-hop IAB network is the tethering-based approach, similar to, e.g., using a smart phone as a WLAN access point. This is equivalent to bundling a terminal and a base station, where the *terminal* part of the IAB node provides access for the *base station* part to the operator internal data network via the CN. The *base station* part can then provide access to the *terminal* part of the subsequent IAB node or to regular UEs, and so on. Such an approach will have minimal impact in terms of standardization and is, in fact, utilized by some commercial LTE solutions on the market providing support for single-hop relaying. However, for the case of multi-hop networks, it will be highly inefficient, because it leads to tunneling-over-tunneling, since a UP tunnel via the CN must be provided for each *terminal* part of the IAB node.

The overhead of tunneling-over-tunneling can be overcome by using the proxying approach of LTE relaying. However, this leads to complex IAB nodes, as the IAB nodes must provide UP/CP gateway/proxying functionality to their child IAB nodes.

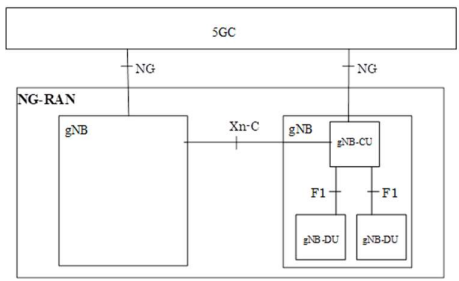

Fig. 2. NR CU/DU split architecture.

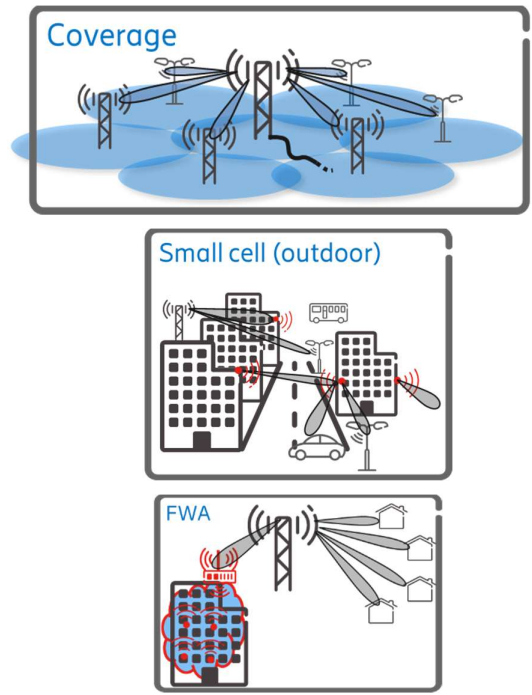

Fig. 3. IAB use case scenarios.

Thus, 3GPP has agreed on a forwarding-based architecture, where each IAB node is assigned an IP address that is routable from a donor base station (and associated L2 addresses), and intermediate IAB nodes forward the packets based on route identifiers/destination addresses. This avoids the tunneling-over-tunneling overhead, eliminates the need for UP/CP gateway/proxy functionality at the IAB nodes and will have minimal impacts on the CN, specially from the UP perspective, as only a limited set of functionalities, such as authentication and authorization, are needed in the CN to support IAB nodes.

Additionally, the CU/DU split architecture described earlier has been chosen for IAB nodes, where the IAB node terminates the DU functionality and a base station (referred to as *IAB-donor*) terminates the CU functionality. Utilizing the CU/DU split architecture will have similar benefits as described for normal gNBs and would also hide the IAB node from external radio access network (RAN) or CN nodes serving the UEs. Centralized mechanisms at the CU, which has an overview of the whole network/path, can be employed for handover decisions, topology change, routing, bearer mapping, etc. As there is no interface between DUs, packets do not need to be forwarded over the air interface between IAB nodes during mobility and dual connectivity operations (e.g. split bearers), reducing the delay and air interface load. Furthermore, the operation and management of a DU will be less complex due to its limited features and capability, compared to a full gNB, which can facilitate the deployment of a denser and more reliable IAB network. One such aspect of flexible deployment is that a DU can be installed at a location that is not necessarily physically secure, without increasing the risk of hacking, as it is the CU that keeps the security information (e.g. keys, ciphering/integrity protection algorithms, etc.).

Fig. 4 shows the UP and CP protocols of multi-hop IAB networks. The IAB node's protocol stack contains two sides, the mobile termination (MT) part, which is used to communicate with a parent node, and a DU part, which is used to communicate with a child node or a normal UE. Also, hop-by-hop RLC is used between the IAB nodes, instead of an E2E RLC between the donor DU and the UE. The main reason for this was that E2E ARQ leads to slower and less efficient retransmissions (since transmission failure at a given hop is not detected until the failure is detected E2E).

### B. Transporting packets over wireless backhaul

Efficient multi-hop forwarding is enabled via the newly-introduced IAB-specific backhaul adaptation protocol (BAP). The IAB-donor assigns a unique L2 address (BAP address) to each IAB node that it controls. In case of multiple paths, multiple route IDs can be associated to each BAP address. The BAP of the origin node (IAB-donor DU for the DL traffic, and the access IAB node for the UL) will add a BAP header to packets they are transmitting, which will include a BAP routing ID (e.g. BAP address of the destination/source IAB node and an optional path ID). Each IAB node will have a routing table (configured by the IAB-donor CU) containing the next hop identifier for each BAP routing ID. Separate routing tables are kept for the DL and UL direction, where the DL table is used by the DU part of the IAB node, while the MT part of the IAB node uses the UL table.

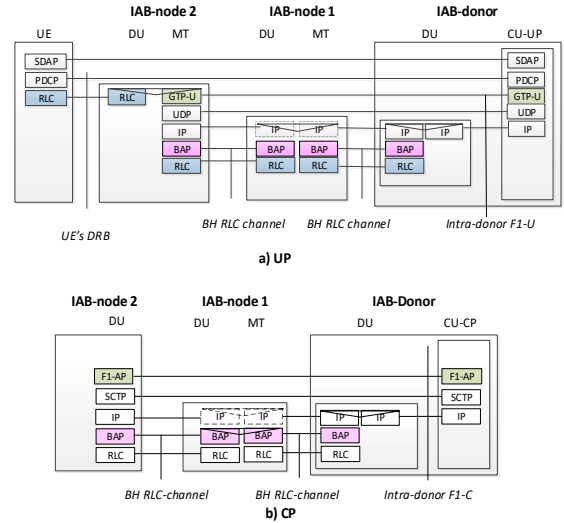

Fig. 4. UP and CP protocols for multi-hop IAB networks.

Backhaul (BH) RLC channels are used for transporting packets between IAB nodes (or between an IAB-donor DU and an IAB node). When it comes to the mapping between UE radio bearers and backhaul RLC channels, two types of mappings are supported, namely N:1 and 1:1 mapping. The N:1 mapping multiplexes several UE radio bearers onto a single BH RLC channel based on specific parameters, such as QoS profile of the bearers. The N:1 is designed for optimal use of BH RLC channels and requires less signaling overhead as a small number of BH RLC channels need to be established. The 1:1 mapping, on the other hand, maps each UE radio bearer onto a separate BH RLC channel, and is designed to ensure fine QoS granularity at UE radio bearer level. 1:1 mapping requires more backhaul RLC channels and more signaling overhead to setup and release BH RLC channels, one for each hop, for each UE radio bearer. Fig. 5 illustrates the routing and bearer mapping, where the UE bearers associated with VoIP and streaming are mapped 1:1, while the bearers for web browsing are mapped N:1.

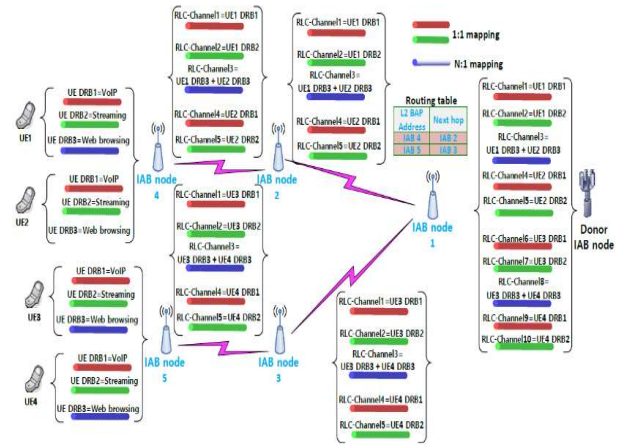

Fig. 5. An illustration of routing and bearer mapping.

## C. Topology adaptation

The wireless backhaul links, due to their usage of mmWave frequencies, could be vulnerable to blockage, e.g., due to moving objects, such as vehicles, seasonal changes (foliage), or infrastructure changes (new buildings). Thus, from a resilience perspective, it is important to ensure that an IAB node can continue to operate (e.g., provide coverage and end-user service continuity) even if an active backhaul path is degraded or lost. For this purpose, 3GPP has agreed on topology adaptation for IAB networks in order to autonomously reconfigure the backhaul network under above-mentioned circumstances. Also, it is desirable to minimize service disruption and packet loss during topology adaptation. IAB topology adaptation can be triggered by the integration of a new IAB node to the topology, detachment/release of an IAB node from the topology, detection of backhaul link overload, deterioration of backhaul link quality, link failure, or other events. The topology adaptation procedure includes three main tasks, namely information collection, topology determination, and topology reconfiguration. For all these tasks, especially for the topology reconfiguration the existing NR Rel-15 procedures for measurements, handover, dual-connectivity, and F1-interface management will serve as baseline.

## D. IAB node intergration into the network

Prior to becoming fully operational, an IAB node executes the IAB integration procedure, shown in Fig. 6. In Step 1, the IAB node connects to the network by using its MT function to execute the initial access procedure for regular UEs. The reuse of the legacy UE initial access procedure is in line with the principle of minimizing the impact on the core network.

In Step 2, the IAB-donor may establish one or more BH RLC channels at one or more intermediate hops towards the newly-joining IAB node (e.g. to setup a certain number of default BH RLC channels between the new IAB node and its parent that could be used for N:1 mapping, BH RLC channel for F1-AP traffic of the DU part of the IAB node, etc.). The IAB-donor also updates the routing tables at intermediate hops, in order to enable the routing towards the IAB node.

In Step 3, the BH RLC connectivity established at Step 2 is used to carry the F1-AP control signaling used to configure the DU function of the IAB node. Once the IAB-DU function has been set up, the IAB node can serve regular UEs, similar to any other DU.

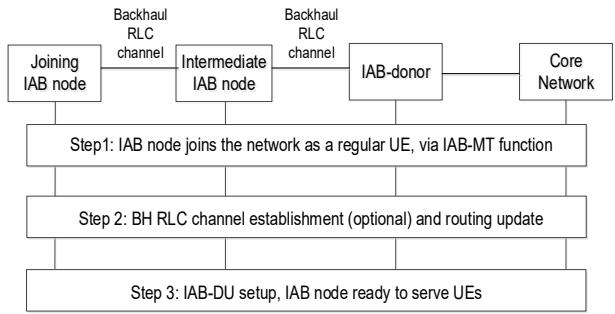

Fig. 6. The steps of IAB node integration procedure.

## V. PHYSICAL LAYER ASPECTS

The IAB backhaul link is based on the NR Rel-15 physical layer with some extensions. IAB nodes support Time Division Multiplexing (TDM), Frequency Division Multiplexing (FDM), and Space Division Multiplexing (SDM) between access and backhaul links at an IAB-node, subject to a half-duplex constraint. However, most of the extensions from NR rel-16 are aimed to enable *in-band* IAB operation, that is, the use of the same carrier frequencies for the MT and DU sides of the IAB node.

In case of in-band operation, the MT part of an IAB node typically cannot receive when the DU part is transmitting and vice versa, also referred to as the IAB *half-duplex constraint*. In this case, there is a need for strict time-domain separation between the IAB-node DU and MT parts. It should be noted though that in some scenarios with high isolation between the DU and the MT, e.g. when the MT and the DU are located on different sides of the same wall, full-duplex IAB operation may be possible.

Similar to NR Rel-15 UEs, MT time-domain resources are configured as *Downlink*, *Uplink*, or *Flexible*, indicating the possible transmission direction of a given resource. However, due to the half-duplex constraint, a certain MT resource configuration does not necessarily imply that the MT is available in the configured transmission direction(s). Rather, this also depends on the configuration of the corresponding DU resource(s).

Similar to the MT, DU time-domain resources are configured as *Downlink*, *Uplink*, or *Flexible*, indicating the allowed transmission direction for a given resource. In order to coordinate the MT and DU and to, for example, handle the half-duplex constraint, DU resources are further configured as *Hard*, *Soft*, or *Not Available*.

A hard DU resource is available in the configured transmission direction(s) without the IAB node having to consider the impact on the corresponding MT resources. In practice, this implies that MT resources corresponding to a hard DU resource (of the same IAB node) are not available, as it cannot be guaranteed that the MT can properly transmit/receive on these resources.

Fig. 7 illustrates how the above principle can be used to create a semi-static resource separation between the DU and MT parts of an IAB node, for a chain of IAB nodes.

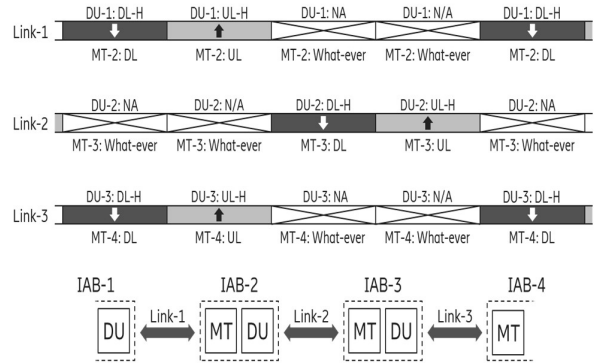

Fig. 7. Semi-static resource separation between DU and MT parts within a chain of IAB nodes.

In contrast to hard resources, a soft DU resource can only be used if that does not impact the MT's ability to transmit and/or receive according to its configuration and scheduling. There are different ways for an IAB node to know that a certain soft DU resource is available, that is, that it can be used without impacting the MT. For example, even though a certain MT resource is configured as uplink, there may be no uplink data available for MT transmission or the MT may not have a valid uplink scheduling grant. In that case, the DU part can use the resource in question.

Furthermore, even if the MT has a valid scheduling grant and there is uplink data available for transmission, the IAB node may know that it is capable of simultaneous DU and MT transmission, in which case the DU can transmit on soft resources. Also, as discussed above, in specific scenarios, an IAB node may be capable of full-duplex between the DU and MT side, in which case the DU can always use a soft resource.

In addition to such *implicit* knowledge of availability of soft DU resources, there is also a possibility for the parent node to provide an *explicit* indication that a certain MT resource will not be used, thus making a corresponding soft DU resource available. Thus, the possibility to configure soft DU resources allows for more dynamic separation between the DU and MT resource allocation. It also allows for the IAB node implementations capable of simultaneous DU and MT operation (transmission and/or reception) to benefit from such capabilities.

Additional IAB extensions to the Rel-15 physical-layer functionality include extended random-access and Synchronization Signal Block (SSB) configurations. The basic principle for IAB-node random access is the same as for NR Rel-15 UEs, i.e. a four-preamble-based random-access procedure [7]. However, it is envisioned that an IAB node may access a (potential) parent node from a larger distance compared to UE access, implying the possible need for a random-access preamble format supporting a larger range. At the same time, IAB-node random access may not be as time-acritical as UE random-access, allowing for longer periodicity, with a corresponding lower overhead for the IAB random-access occasions. To enable this, there is a possibility for separate Random Access Channel (RACH) configurations, using different preamble formats and different random-access-occasion periodicities, for UEs and IAB nodes. To avoid collisions between random-access occasions at the DU and MT sides, the random-access occasions for IAB nodes can be configured with an additional time-domain offset.

A similar approach is taken to avoid collision between DU SSB transmissions and SSB discovery/measurements done by the corresponding MT. In the typical case, the half-duplex constraint prevents an IAB node to search for and measure on SSB transmissions of other nodes at the same time as it is transmitting SSB for UE cell search. Thus, for IAB nodes, the Rel-15 SSB transmission configurations are extended to enable SSB transmission with additional time-domain offsets. Such time-offset SSB transmissions do not comply with Rel-15 and can thus not be used for UE cell search without breaking backwards compatibility. However, they can be used to enable IAB inter-node discovery and measurements.

With regard to the transmission timing of IAB nodes, at least when operating in unpaired (TDD) spectrum, base stations should transmit in a synchronous way with time alignment between transmissions. One way to achieve such inter-node time alignment is to rely on location information (e.g. Global Positioning System (GPS)) reception at all IAB nodes. However, IAB also allows for over-the-air (OTA) synchronization where a given IAB node derives its transmit timing from downlink signal received from its parent node, which may be another IAB node or the IAB donor node. In order for this to be possible, the IAB node needs an estimate of the propagation delay on the link from the parent node. This can be derived from the knowledge about the timing offset between the DL reception and UL transmission at the IAB node (inherently known by the IAB node) and the corresponding offset between the DL transmission timing and UL reception timing at the parent node (explicitly provided by the parent node to the IAB node).

## V. CONCLUSIONS

The main aim of IAB networks is to facilitate the dense deployment of 5G networks at mmWave frequencies without the need for a fiber connection to each base station. By utilizing part of the large spectrum available at mmWave frequencies for wireless backhauling, IAB will considerably reduce the deployment cost of 5G networks, while providing a comparable performance to fiber deployment. In this paper, we presented an overview of the multi-hop IAB network currently being standardized in 3GPP rel-16. We provided the key design principles and features of IAB in 3GPP, followed by an overview of the architecture, protocol and physical layer aspects, and the rationale behind the main design choices. For a more in-depth overview IAB networks, the reader is referred to the 3GPP technical report on IAB [8].

REFERENCES

[1] Ericsson, "Ericsson Mobility Report", https://www.ericsson.com/assets/local/mobility-report/documents/2019/ericsson-mobility-report-june-2019.pdf , June 2019

[2] H. A. Willebrand and B. S. Ghuman, "Fiber optics without fiber," IEEE Spectrum, vol. 38, no. 8, pp. 40–45, Aug. 2001.

[3] C. Dehos, J. L. González, A. D. Domenico, D. Kténas and L. Dussopt, "Millimeter-wave access and backhauling: the solution to the exponential data traffic increase in 5G mobile communications systems?," IEEE Commun. Mag., vol. 52, no. 9, pp. 88-95, Sept. 2014.

[4] Y. Li, E. Pateromichelakis, N. Vucic, J. Luo, W. Xu and G. Caire, "Radio Resource Management Considerations for 5G Millimeter Wave Backhaul and Access Networks," IEEE Commun. Mag., vol. 55, no. 6, pp. 86-92, June 2017.

[5] 3GPP, "Overview of 3GPP Release 10 V0.2.1," Available at: https://www.3gpp.org/ftp/Information/WORK_PLAN/Description_Releases/

[6] 3GPP, RP-182882, New WID: Integrated Access and Backhaul for NR, http://www.3gpp.org/ftp/TSG_RAN/TSG_RAN/TSGR_82/Docs/RP-182882.zip, Dec 2018.

[7] E. Dahlman, S. Parkvall, J. Sköld, "5G NR - The Next Generation Wireless Access Technology", Academic Press, Oxford, UK, 2018.

[8] 3GPP, "Study on Integrated Access and Backhaul", TR 38.874, http://www.3gpp.org/ftp//Specs/archive/38_series/38.874/38874-g00.zip, Jan 2019